\documentclass[3p,times,sort&compress]{elsarticle}
\newif\ifsubmitted
\submittedfalse

\ifsubmitted
\usepackage{ecrc}

\volume{00}

\firstpage{1}

\journalname{Discrete Mathematics}

\runauth{J.~Ne\v set\v ril, P.~Ossona de Mendez, S.~Siebertz}

\jid{dm}

\jnltitlelogo{Discrete Maths}

\CopyrightLine{2022}{This work is licensed under a Creative Commons Attribution 4.0 International License.}
\else
\nopreprintlinetrue
\fi

\usepackage{amssymb}

\usepackage{amsmath}
\usepackage{cleveref}
\usepackage{mathrsfs}
\usepackage{accents}
\usepackage{complexity}
\usepackage{microtype}
\usepackage[all,defaultlines=3]{nowidow}
\usepackage{xcolor}
\usepackage[all]{xy}
\newclass{\FOM}{FO\mathord{+}MOD}
\newclass{\CMSO}{CMSO}
\newclass{\MSO}{MSO}
\newclass{\MSOE}{MSO_2}
\newcommand{\Cc}{\ensuremath{\mathscr C}}
\newcommand{\Dd}{\ensuremath{\mathscr D}}

\DeclareMathOperator{\rank}{rank}
\DeclareMathOperator{\srank}{set-rank}
\DeclareMathOperator{\Supp}{Support}
\DeclareMathOperator{\Gaif}{Gaifman}
\DeclareMathOperator{\tp}{Tp}
\newcommand{\mexists}[2]{\exists^{{\hskip .05em}{#1}[{#2}]}}

\newtheorem{theorem}{Theorem}
\newtheorem{lemma}{Lemma}

\newtheorem{corollary}{Corollary}

\newtheorem{definition}{Definition}

\newtheorem{conjecture}{Conjecture}
\newproof{proof}{Proof}

\crefname{corollary}{Corollary}{Corollaries}
\begin{document}

\begin{frontmatter}


\ifsubmitted
\dochead{CSGT 2022 special issue}
\fi

\newcommand{\ERCagreement}{
		\begin{minipage}{.6\textwidth}
			\footnotesize
			This paper is part of projects that have received funding from the European Research Council (ERC) under the European Union's Horizon 2020 research and innovation programme (grant agreements No 810115 -- {\sc Dynasnet} and 948057 -- {\sc BOBR}) and from the German
			Research Foundation (DFG) with grant agreement
			No 444419611.
		\end{minipage}\hfill
		\begin{minipage}{.34\textwidth}
			\hfill\includegraphics[height=1cm]{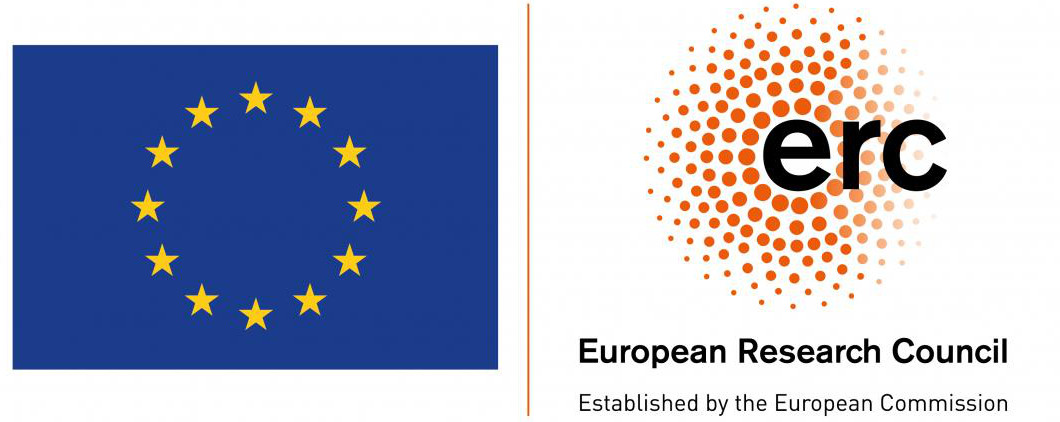}
			\hfill\includegraphics[height=1cm]{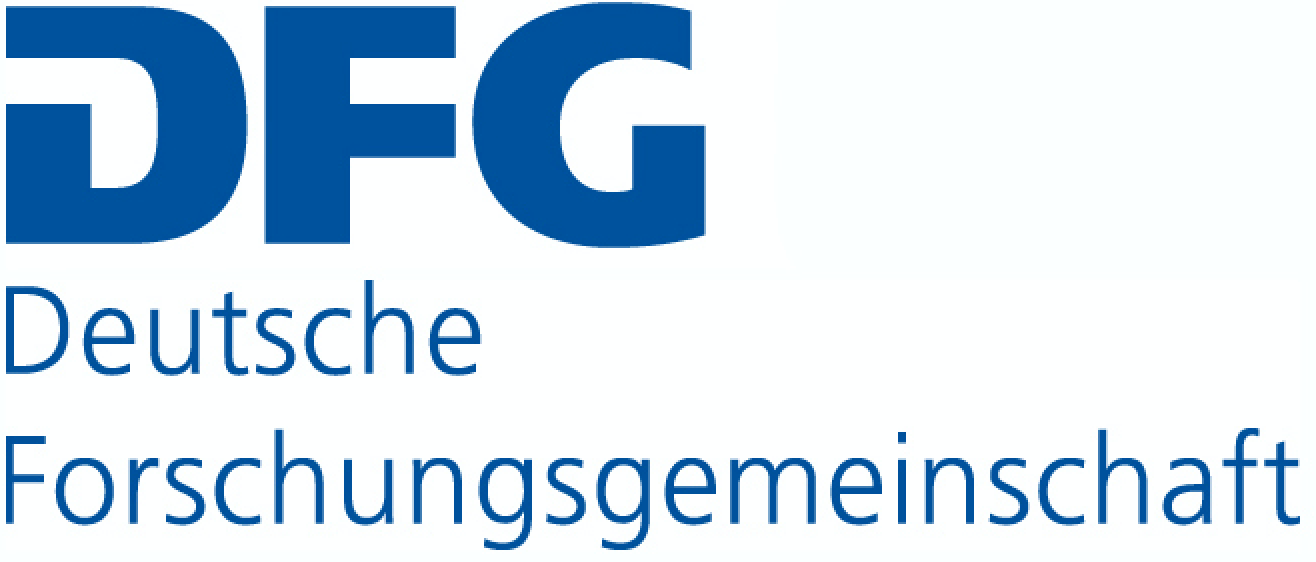}
		\end{minipage}
}
\title{Modulo-Counting First-Order Logic on Bounded Expansion Classes\tnotemark[1]}
 \tnotetext[1]{\ERCagreement}
\author[iuuk]{Jaroslav Ne\v set\v ril}
\address[iuuk]{Computer Science Institute of Charles University (IUUK), Praha, Czech Republic}
\ead{nesetril@iuuk.mff.cuni.cz}
\author[cams,iuuk]{Patrice Ossona de Mendez}
\address[cams]{Centre d'Analyse et de Math\'ematiques Sociales (CNRS, UMR 8557),
	Paris, France}
\ead{pom@ehess.fr}
\author[brem]{Sebastian Siebertz}
\address[brem]{University of Bremen, Bremen, Germany}
\ead{siebertz@uni-bremen.de}
\begin{abstract}
We prove that, on bounded expansion classes, every  first-order formula with modulo counting is equivalent, in a linear-time computable monadic expansion,  to an existential first-order formula.

As a consequence, we derive, on  bounded expansion classes, that first-order transductions with modulo counting have the same encoding power as existential first-order transductions.
Also, modulo-counting first-order model checking and computation of the size of sets definable in modulo-counting first-order  logic can be achieved in linear time on bounded expansion classes.

As an application, we prove that a class has structurally bounded expansion if and only if it is a class of bounded depth vertex-minors of graphs in a bounded expansion class.

We also show how our results can be used to implement fast matrix calculus on bounded expansion matrices over a finite field.
\end{abstract}

\begin{keyword}
	Graph theory \sep first-order logic \sep bounded expansion class \sep modulo counting \sep transduction \sep model checking \sep vertex minor \sep matrix calculus
	
	\MSC[2020] 05C99, 03B70, 03C10, 68R10



\end{keyword}

\end{frontmatter}


\section{Introduction}
\label{sec:intro}
\emph{Algorithmic meta-theorems} commonly refer to results proving that every computational problem definable in a given logic can be solved efficiently on every class of graphs satisfying certain conditions. A prototypical example is the celebrated theorem of Courcelle~\cite{Courcelle1} stating that every graph property definable in {\MSOE} (Monadic Second-Order logic with quantification over sets of vertices or edges) can be decided in linear time on all graph classes of bounded tree width. Further work established meta-theorems for other logics. For instance, Courcelle, Makowski, and Rotics~\cite{Courcelle2000a} proved that {\MSO} model-checking (Monadic Second Order logic with quantification over sets of vertices) is fixed-parameter tractable on classes with bounded cliquewidth.

For the much less expressive first-order logic (\FO), fixed parameter tractability on  graph classes of bounded maximum degree was established by Seese~\cite{Seese1996}. This result was further extended to classes with bounded local tree-width~\cite{Frick2004} and to classes
excluding a minor~\cite{Flum2001}, to classes locally excluding a minor~\cite{local_minor}, to classes with bounded expansion~\cite{DKT,DKT2}, and eventually to nowhere dense classes~\cite{Grohe2013}, which are the most general monotone classes of graphs on which {\FO} model checking is not \AW[*]-complete. Tractability results of {\FO} model-checking for (non-sparse) hereditary classes of graphs have been proved since, for map graphs~\cite{FO_map_graphs} and, more generally, for structurally bounded expansion classes (i.e.\ to transductions of classes with bounded expansion)~\cite{SBE_drops,SBE_TOCL}, and for classes with bounded twin-width~\cite{twin-width1}. We remark that for the last two results additional decompositions need to be given with the input. In this setting, it has been conjectured~\cite{gajarskynew} that \FO-model checking is fixed-parameter tractable on every monadically NIP class of graphs.

It is well-known that first-order logic lacks the power to count, which led to the study of extensions of first-order logic with counting mechanisms.
The language {\FOM} is an extension of first-order logic {\FO} with modulo-counting quantifiers of the for
$\mexists{a}{b}x\enspace\psi$, expressing that the number of witnesses $x$ that satisfy~$\psi$ is congruent to $a$ modulo $b$. For example, we can express that a graph contains $k$ vertices $v_1,\dots,v_k$ such that every two of them have an odd number of common neighbors (which is not \FO-expressible).

It has been proved that {\FOM} model-checking is fixed parameter tractable over classes of graphs with bounded degree~\cite{berkholz2018answering} and over classes of graphs with bounded twin-width~\cite{bonnet2022twin}. In this paper, we extend the former by proving that {\FOM} model-checking is fixed parameter tractable over every class with bounded expansion
(\Cref{cor:fpt}), by proving a form of quantifier elimination for {\FOM} on bounded expansion classes (\Cref{thm:main,cor:main2}).

A \emph{transduction} $\mathsf T$  is an operation that allows to define a set $\mathsf T(G)$ of graphs (or structures) from a single graph (or structure) $G$. It basically proceeds in two steps\footnote{Note that sometimes an additional \emph{copy step}, which consists of a finite blowing of the vertices, has also been considered in the literature.}, namely a \emph{coloring step}, in which the vertices of the graph are ``colored'' in all the possible ways by adding a number of unary predicates; then, to each colored graph is associated a new graph, whose vertex set and edge set are both defined using logical formulas (which can use the coloring predicates and formally define the transduction).   If $\Cc$ is a set of graphs, the set of all the graphs that can be obtained from graphs in $\Cc$ using the transduction $\mathsf T$ is $\mathsf T(\Cc)=\bigcup_{G\in\Cc}\mathsf T(G)$, and we say that a class $\Dd$ is a $\mathsf T$-transduction of a class $\Cc$ if $\Dd\subseteq\mathsf T(\Cc)$. Intuitively, that $\Dd$ is a $\mathsf T$-transduction of $\Cc$ means that every graph in $\Dd$ can be encoded in a graph $\Cc$ by some vertex coloring, which can be decoded using the two formulas defining $\mathsf T$. Existence of a transduction allowing to encode a class $\Dd$ in a class $\Cc$ defines a very rich (bounded semi-lattice) quasi-order~\cite{arboretum,Tquasiorder}.

While it is obvious that {\FOM} is strictly more expressive than {\FO} --- it allows, for instance, to express that the domain has even size ---  it is not clear whether {\FOM} transductions are more powerful than {\FO} transductions. In this paper, we prove that this is not the case when we restrict to structurally bounded expansion classes. Precisely, we prove that for every structurally bounded expansion class $\Cc$ and every {\FOM} transduction $\mathsf T$, there exists an {\FO} transduction $\widehat{\mathsf T}$ such that, for every graph $G\in\Cc$, the set $\widehat{\mathsf T}(G)$ of all graphs produced by $\widehat{\mathsf T}$ from $G$ includes the set $\mathsf T(G)$ of all graphs produced by $\mathsf T$ from~$G$. We deduce, for instance, that the class of depth-$d$ shallow vertex minors of the graphs in a structurally bounded expansion class $\Cc$ has itself structurally bounded expansion. Moreover, when we restrict to bounded expansion classes, the transduction $\widehat{\mathsf T}$  can be required to be an existential transduction (\Cref{cor:trans}).

As an example, we consider local complementation of vertices, which consists of the complementation in the neighborhood of  the selected vertices \cite{bouchet1987isotropic}, and a shallow version of vertex minors.

The study of vertex minors recently gained much attention, as it appeared to be a good analog of the notion of minors in the study of hereditary classes of (possibly dense) graphs, possibly allowing an analog of the celebrated Structure Theorem of Robertson and Seymour for classes excluding a graph as a vertex minor (See \cite{mccarty2021local}).
As the study of shallow minors appeared to be of particular interest, and leading to a taxonomy of sparse graph classes \cite{Sparsity}, it looks natural to consider a shallow version of vertex minors.
It is not too difficult to prove that, on a class with bounded expansion, one can obtain  bounded depth minors via an \FO-transduction, and bounded depth vertex minors via an \FOM-transduction. The fact that \FOM-transductions and \FO-transductions coincide on bounded expansion classes is an intriguing consequence of our results. For instance, combined with results of \cite{SBE_TOCL}, it implies that a class is an \FO-transduction of a class with bounded expansion if and only if it is a class of bounded depth vertex minors of a class with bounded expansion (\Cref{cor:vm}).

We also consider some application of these results to matrix calculus, when the considered matrices have coefficients in a finite ring (essentially $\mathbb F_p$), and when the non-zero entries of all the considered matrices (but some constant matrices with low ``set-rank'', like the all-$1$ matrix $J$) are together supported by the adjacency matrix of a  graph in a fixed bounded expansion class (\Cref{thm:mat-calc}).

\section{Preliminaries}
\subsection{Basics from Model Theory}
In this paper, we consider finite graphs, as well as more general structures. Recall that a \emph{signature} $\sigma$ is a set of relation and function symbols with arities. A \emph{$\sigma$-structure} $\mathbf M$ has a finite set $M$ as its \emph{domain} and an interpretation of the relation and function symbols: if $R$ is a $k$-ary relation symbol, then we denote by $R^{\mathbf M}$ its interpretation in $\mathbf M$, so that, for every tuple $\bar v\in M^k$ we have
$R^{\mathbf M}(\bar v)$ if and only if $\mathbf M\models R(\bar v)$. Similarly, if $f$ is a $k$-ary function symbol, $u=f^{\mathbf M}(\bar v)$ if and only if $\mathbf M\models (u=f(\bar v))$.

 Let $\sigma$ be such a signature,
let $\mathbf M$ be a $\sigma$-structure, and let $X$ be a subset of its domain $M$. We denote by $\mathbf M[X]$ the \emph{restriction} of $\mathbf M$ to $X$, which is a $\sigma$-structure with domain $X$, in which
$R^{\mathbf M[X]}(\bar v)$ is equivalent to  $R^{\mathbf M}(\bar v)$ for every tuple $\bar v$ of elements of $X$, and where, for every $v\in X$, we have
\[
f^{\mathbf M[X]}(v)=\begin{cases}
	f^{\mathbf M}(v)&\text{if }f^{\mathbf M}(\bar v)\in X\\
	v&\text{otherwise.}
\end{cases}
\]

The \emph{Gaifman graph} $\Gaif(\mathbf M)$ of a $\sigma$-structure is the graph with vertex set $M$ where $u\neq v$ are adjacent if they belong jointly to some relation $R^{\mathbf M}$ (with $R\in\sigma$) or if one is the image of the other by some function $f^{\mathbf M}$ (with $f\in\sigma$).

An \emph{expansion}  (or \emph{lift}) of a $\sigma$-structure $\mathbf M$ is a $\sigma'$-structure $\mathbf M^+$ with the same domain, where $\sigma'\supset\sigma$ and all the relations and functions in $\sigma$ have the same interpretation in $\mathbf M$ and $\mathbf M^+$. 
If $\sigma'\setminus\sigma$ consists only of unary relation symbols, then $\mathbf M^+$ is a \emph{monadic expansion} of $\mathbf M$.
If $\mathbf M^+$ is an expansion of $\mathbf M$, then $\mathbf M$ is a \emph{reduct} of $\mathbf M^+$.

We denote by $\FO[\sigma]$ the first-order logic in the language of $\sigma$-structures, and by $\FOM[\sigma]$ its extension containing modulo counting quantifiers $\mexists{a}{b}$, expressing the existence of $a$ modulo $b$ elements. 

Let $\sigma,\sigma'$ be finite signatures.
A \emph{simple \FO-interpretation}\footnote{Note that the term ``interpretation'' has a meaning deeply different than above, where it meant something like ``instanciation".} $\mathsf I$ of $\sigma'$-structures in $\sigma$-structure consists of a formula~$\phi_R$ (resp.\ $\phi_f$) in $\FO[\sigma]$ for each relation symbol $R\in\sigma'$ (resp.\ for each function symbol $f\in\sigma'$), as well as formula $\nu\in\FO[\sigma]$ with a single free variable. If $\mathbf M$ is a $\sigma$-structure, the $\sigma'$-structure $\mathsf I(\mathbf M)$ has domain $\nu(\mathbf M)=\{v\mid \mathbf M\models\nu(v)\}$ and is such that, for every $u\in \nu(\mathbf M)$ and every $\bar v\in \nu(\mathbf M)^k$, 
\begin{itemize}
	\item if $R\in\sigma'$ is a relation symbol with arity $k$, then $\phi_R$ has $k$ free variables and 
	\[\mathsf I(\mathbf M)\models R(\bar v)\quad\iff \quad\mathbf M\models \phi_R(\bar v);\]
		\item if $f\in\sigma'$ is a function symbol with arity $k$, then $\phi_f$ has $k+1$ free variables and 
	\[\mathsf I(\mathbf M)\models (u=f(\bar v))\quad\iff \quad\mathbf M\models \phi_f(u,\bar v).\]
\end{itemize}

When defining an interpretation, it will be convenient to omit any trivial formula defining identically a relation or a function
(such as $\phi_R(\bar x):=R(\bar x)$ with $R\in\sigma\cap\sigma'$ or $\phi_f(y,\bar x):=(y=f(\bar x))$ with $f\in\sigma\cap\sigma'$). Hence, for the sake of readability, any interpretation $\mathsf I$ of $\sigma'$-structures in $\sigma$-structure is also considered as an interpretation of $\hat\sigma'$-structures in $\hat\sigma$-structures if $\sigma\subseteq \hat\sigma$ and $\hat\sigma'\setminus\sigma'\subseteq \hat\sigma$.
A \emph{basic interpretation} is a simple interpretation where $\nu$ is a tautology. An interpretation is \emph{quantifier-free} if all the formulas defining it are quantifier-free.

Let $\mathsf I$ be a simple interpretation of $\sigma'$-structures in $\sigma$-structures. Then, $\mathsf I$ 
defines a natural mapping $\mathsf I^\ast:\FO[\sigma']\rightarrow\FO[\sigma]$ such that 
(for a formula $\phi$ with $k$ free variables and a tuple $\bar v$ of $k$ elements of the domain of $\mathsf I(\mathbf M)$ we have
\[
\mathsf I(\mathbf M)\models \phi(\bar v)\quad\iff\quad
\mathbf M\models \mathsf I^\ast(\phi)(\bar v).
\]

Moreover, if $\mathsf I$ is quantifier-free, then $\mathsf I^\ast$ maps existential formulas of $\FO[\sigma']$ to existential formulas of $\FO[\sigma]$. (Note that, in general, 
$\mathsf I^\ast$  does not map quantifier-free formulas to quantifier-free formulas because of the possibility to define new functions.)

\pagebreak
In this paper, we consider two types of structures:
\begin{itemize}
	\item \emph{Colored graphs} are $\sigma$-structures, where $\sigma$ consists of a single binary relation $E$ and some number of unary relations, with the property that
	$E$ is symmetric ($\forall x,y\ E(x,y)\leftrightarrow E(y,x)$) and anti-reflexive ($\forall x\ \neg E(x,x)$).
	\item \emph{Guided pointer structures} are $\sigma'$-structures, where $\sigma'$ consists of a single binary relation $E$, some number of unary relations, and some unary functions, with the property that $E$ is symmetric  and anti-reflexive, and that  every function $f\in\sigma'$ is \emph{guided}, meaning 
	$\forall x,y\ (y=f(x))\rightarrow ((x=y)\vee E(x,y))$.
\end{itemize}

A \emph{guided expansion} of a colored graph $G$ (with signature $\sigma$) is a guided pointer structure $\mathbf M$ (with signature $\sigma'\supseteq\sigma$), which is an expansion of $G$. 
As  the adjacency relation $E$ belongs to $\sigma$,  the structures $G$ and $\mathbf M$ have the same adjacency relation, hence the same Gaifman graph.

\subsection{Graph Theory}

Low tree-depth decompositions have been introduced in~\cite{Taxi_tdepth} and used in~\cite{POMNI,POMNII,POMNIII} to characterize bounded expansion classes and prove some structural and algorithmic properties of them. The interested reader is referred to~\cite{japan04,Sparsity} for an in-depth study of these decompositions.

\begin{definition}
	The {\em tree-depth} ${\rm td}(G)$ of a graph $G$ is
	defined as
	the minimum height of a rooted forest $Y$ such that $G$ is a subgraph of the closure of $Y$ (that is of the graph obtained by adding edges between a vertex and all its ancestors). In particular, the tree-depth of a disconnected graph is the maximum of the tree-depths of its connected components.
\end{definition}

\begin{definition}
	A {\em low tree-depth decomposition} with parameter $p$ of a graph $G$ is a coloring of the vertices of $G$, such that any subset $I$ of at most $p$ colors induces a subgraph with tree-depth at most $|I|$. The minimum number of colors in a low tree-depth decomposition with parameter $p$ of $G$ is denoted by $\chi_p(G)$.
\end{definition}

\begin{definition}
	\label{def:BE}
	A class $\mathscr{C}$ has {\em bounded expansion} if there exists a function $f:\mathbb{N}\rightarrow\mathbb{N}$ such that every topological minor~$H$ of a graph $G\in\mathscr{C}$ has an average degree bounded by $f(p)$, where $p$ is the maximum number of subdivisions per edge needed to turn $H$ into a subgraph of $G$.
\end{definition}

Existence of bounded size low tree-depth decompositions characterizes classes with bounded expansion:

\begin{theorem}[\cite{POMNI}]
	\label{thm:chiBE}
	Let $\mathscr{C}$ be a class of graphs, then the following are equivalent:
	\begin{enumerate}
			\item for every integer $p$ we have $\sup_{G\in\mathscr{C}}\chi_p(G)<\infty$;
			\item the class $\mathscr{C}$ has bounded expansion.
		\end{enumerate}
\end{theorem}

A slightly stronger notion than low tree-depth decomposition is the notion of $p$-centered coloring. A \emph{$p$-centered coloring} of a graph $G$ is a coloring $\gamma$ of the vertices of $G$ with the property that for every connected (induced) subgraph of $H$, either one color appears exactly once in the vertices of $H$, or $H$ gets at least $p$ colors. It is easily checked that every $(p+1)$-centered coloring defines a low tree-depth decomposition with parameter $p$ \cite{Taxi_tdepth}. A \emph{centered coloring} of a graph $G$ is a coloring $\gamma$ of the vertices of $G$ with the property that for every connected subgraph of $H$,  one color appears exactly once in the vertices of $H$. (Thus, a centered coloring is an $\infty$-centered coloring.) Note that if $\gamma$ is a $(p+1)$-centered coloring of a graph $G$ and $H$ is a subgraph of $G$ that gets at most $p$ colors, then the restriction of $\gamma$ to~$V(H)$ is a centered coloring of $H$.

From an algorithmic point of view, an important property is that a (non-optimal)  bounded size $p$-centered coloring can be computed in linear time on every fixed class with bounded expansion.

\begin{theorem}[\cite{POMNII}]
	\label{thm:chiBE_alg}
	Let $\mathscr{C}$ be a class of graphs with bounded expansion. Then, there is a function $B:\mathbb N\rightarrow\mathbb N$ (which depends on $\mathscr C$) and an  algorithm (independent of $\mathscr C$)
	that computes, for $G\in\mathscr{C}$ and $p\in\mathbb N$, a $p$-centered coloring 
	of $G$ with at most $B(p)$ colors in time $O_{B(p),p}(|G|)$.
\end{theorem}

Low tree-depth decompositions are the central tool of the \FO-model checking algorithm presented in~\cite{DKT,DKT2}. A main result of these papers is the following, which we formulate in our setting.

\begin{theorem}[\cite{DKT,DKT2}]
	\label{thm:DKT}
	For every \FO-formula $\varphi(\bar x)$ and every bounded expansion class $\Cc$ of colored graphs, there exists a quantifier-free formula $\tilde\phi$ and a 
	linear time computable map $\Lambda$ such that, for every $G\in\mathscr C$, $\Lambda(G)$ is a guided expansion of $G$ such that
	\[
	G\models \varphi(\bar v)\quad\iff\quad \Lambda(G)\models\tilde\varphi(\bar v).
	\]
\end{theorem}

As an easy corollary, one gets

\begin{corollary}[\cite{DKT,DKT2}]
	\label{cor:DKT}
	For every \FO-formula $\varphi(\bar x)$ and every bounded expansion class $\Cc$ of colored graphs, there exists an existential formula $\hat\phi$ and a 
	linear time computable map $\hat\Lambda$ such that, for every $G\in\mathscr C$, $\hat\Lambda(G)$ is a monadic expansion of $G$ such that
	\[
	G\models \varphi(\bar v)\quad\iff\quad \hat\Lambda(G)\models\hat\varphi(\bar v).
	\]
\end{corollary}

This theorem allows, for instance, to count the number of satisfying assignments of a first-order formula in linear time, or to enumerate these satisfying assignments in constant time per assignment (after a linear time preprocessing phase) by reducing the problem to the case where the formula is an existential formula \cite{Kazana2013}.

\section{Elimination of modulo counting quantifiers}
Our main result is an extension of \Cref{thm:DKT,cor:DKT} to {\FOM} formulas.

\begin{theorem}
	\label{thm:main}
	Let $\mathscr C$ be a bounded expansion class of colored graphs with guided
	pointers (with signature $\sigma$) and let $\varphi\in\FOM[\sigma]$.
	Then, there exists a bounded expansion class $\mathscr C'$ of guided pointer structures with signature $\sigma'\supseteq\sigma$, a linear-time computable map $L:\mathscr C\rightarrow\mathscr C'$, and a quantifier-free formula $q\in\FO[\sigma']$ with the following properties:
	\begin{itemize}
		\item $L$ maps each $\mathbf M\in\mathscr C$ to some guided expansion $L(\mathbf M)\in\mathscr C'$;
		\item for every $\mathbf M\in\mathscr C$ and every
		$\bar v\in M^{|\bar x|}$, we have
\[
	\mathbf M\models \varphi(\bar v)\quad\iff\quad L(\mathbf M)\models q(\bar v).
\]
	\end{itemize}
\end{theorem}

The following corollary is a direct consequence of \Cref{thm:main,cor:DKT}.
\begin{corollary}
	\label{cor:main2}
		Let $\mathscr C$ be a bounded expansion class of colored graphs (with signature $\sigma$) and let $\varphi\in\FOM[\sigma]$.
	Then, there exists a bounded expansion class $\mathscr {\hat C}$ of colored graphs with signature $\hat\sigma\supseteq\sigma$, a linear-time computable map $\hat L:\mathscr C\rightarrow\mathscr{\hat C}$, and an existential formula $\hat\varphi\in\FO[\sigma']$ with the following properties:
	\begin{itemize}
		\item $\hat L$ maps each $G\in\mathscr C$ to some monadic expansion $\hat L(G)\in\mathscr{\hat C}$;
		\item for every $G\in\mathscr C$ and every
		$\bar v\in V(G)^{|\bar x|}$, we have
		\[
		G\models \varphi(\bar v)\quad\iff\quad \hat L(\G)\models \hat\varphi(\bar v).
		\]
	\end{itemize}
\end{corollary}
\Cref{cor:main2} and \cite{DKT,Kazana2013} directly imply the following.
\begin{corollary}
	\label{cor:fpt}
	Let $\mathscr C$ be a class with bounded expansion.
	
	Then, on $\mathscr C$, \FOM-model checking is fixed parameter linear-time  and, more generally,  \FOM-definable sets can have their cardinality computed in linear-time  and be enumerated with constant delay after a linear-time preprocessing.
\end{corollary}

Our last corollary shows that the possibility to use modulo counting does not increase the expressive power of transduction on structurally bounded expansion classes.

\begin{corollary}
	\label{cor:trans}
	Every \FOM-transduction on a class of graphs with  bounded expansion is subsumed by an existential \FO-transduction.
\end{corollary}

\subsection{Overview of the proof}
\label{sec:overview}
The proof of \Cref{thm:main} will proceed by induction, eliminating one modulo counting quantifier at a time.
The induction step involves the introduction of several intermediate structures, related as follows (note that the direction of arrows indicates the direction of encoding, the proof goes clockwise from $\mathbf{M}$ to $L(\mathbf{M})$):
\[
\xy\xymatrix@C=3cm{
	\mathbf M\ar@{-->}^{\text{$p$-centered coloring}}[r]&\mathbf M^+\ar@{..>}^{\text{induced subgraphs}}[r]&\mathbf M^+_{\bar t,t'}&Y^+_{\bar t,t'}\ar_{\text{interpretation }\mathsf I_S}[l]\ar@{-->}[d]^{\parbox{2.1cm}{\flushleft\scriptsize modulo counting\\ quantifier elimination\\~}}\\
	L(\mathbf M)&\mathbf M^\ast\ar@{-->}^{\text{quantifier elimination}}[l]\ar@{..>}_{\text{induced subgraphs}}[r]&\mathbf M^\ast_{\bar t,t'}\ar_{\text{interpretation }\mathsf I_Y}[r]&Y^\ast_{\bar t,t'}
}
\endxy
\]

Starting from a guided pointer structure $\mathbf M$, we first compute a $p$-centered coloring of its Gaifman graph, for a value~$p$ depending on the considered formula, thus obtaining a monadic expansion $\mathbf M^+$ of $\mathbf M$.
Then, we prove that the formula decomposes into sub-formulas, which can be evaluated on substructures $\mathbf M_{\bar t,t'}$ induced by at most $p$ colors. These substructures can be interpreted in bounded height rooted colored forests $Y^+_{\bar t,t'}$. Then, we can use the strong structural properties of bounded height forests to eliminate the modulo counting quantifier in a monadic expansion~$Y^\ast_{\bar t,t'}$ of $Y^+_{\bar t,t'}$. 
The obtained first-order formula is transferred back to the induced substructures (after adapted monadic expansion $\mathbf M^\ast_{\bar t,t'}$) and, eventually, to a monadic expansion $\mathbf M^\ast$ of the original structure. At this point, we are left with a standard first-order formula, and we can invoke \Cref{thm:DKT} to eliminate all first-order quantifiers, thus ending with a quantifier-free formula to be evaluated in a monadic expansion $L(\mathbf M)$ of the original structure $\mathbf M$.

In order to prove \Cref{thm:main}, we need to first introduce some preliminary constructions.

\subsection{The transductions $\mathsf I_Y$ and $\mathsf I_S$}
\label{sec:Y}
We start by introducing two interpretations, which allow encoding and decoding guided pointer structures with Gaifman graph of bounded tree-depth in colored rooted forests with bounded height. These are reminiscent of the transductions introduced in \cite{arboretum} for encoding a structure in its Gaifman graph.

Consider a guided pointer structure $\mathbf M$ with signature $\sigma'$ and a centered coloring 
$\gamma:\Gaif(\mathbf M)\rightarrow X$ of the Gaifman graph of $\mathbf M$ with colors in a finite subset $X$ of $\mathbb N$ with $|X|=h$, witnessing ${\rm td}(\Gaif(\mathbf M))\leq h$.

We define the monadic expansion $\mathbf M^+$ of $\mathbf M$ inductively as follows: start with $i=1$ and $G_1=\Gaif(\mathbf M)$. For each connected component $H$ of $G_i$, we mark by $P_i$ in $\mathbf M$ a uniquely colored vertex of $H$.
Then we  define $G_{i+1}$ as the graph obtained from $G_i$, by deleting all the vertices marked $P_i$ in $\mathbf M$, and we increase $i$. Note that $G_{h+1}$ is empty, as every connected component of $G_i$ gets at most $h+1-i$ colors.
This process defines a rooted forest with height at most~$h$, which contains $\Gaif(\mathbf M)$ in its closure (See, for instance, \cite{Sparsity}).

As graphs with tree-depth at most $h$ do not contain paths of length at least $2^h$ as subgraphs, we can define a formula~$\varpi(x,y)$ defining the parent function of the rooted forest constructed from the centered coloring.
This can be done using the  characteristic property that $y$ is the parent of $x$ if either $x$ is marked by $P_1$ and $x=y$ (meaning that $x$ is a root), or $x$ is marked by some $P_i$ with $1<i\leq h$, $y$ is marked by $P_{i-1}$, and there is a path with length at most $2^h$ linking~$x$ and $y$ and no internal vertex marked $P_j$ with $j<i$.  It is easily checked that $\varpi(x,y)$ defines the parent function~$\pi$ of a rooted forest (meaning that  $\pi(x)=y$ if and only if $\varpi(x,y)$).

We define the interpretation $\mathsf I_Y$ as follows: 
two vertices $x$ and $y$ are adjacent if and only if $\varpi(x,y)\vee \varpi(y,x)$ (and we keep all the unary relations in $\mathbf M^+$). 
The adjacency relation $E$ and the functions defined on $\mathbf M$ will be encoded by means of predicates $T_{E,i,j}$ and $T_{f,i,j,\epsilon}$ by taking advantage of the fact that they relate pairs of vertices comparable in the rooted forest.
Precisely, for each $1\leq j<i\leq h$ we add a predicate $T_{E,j,i}(v)$ expressing that $v$ is marked $P_i$ and is adjacent in $\mathbf M$ to a vertex marked $P_j$, as well as predicates $T_{f,j,i,\epsilon}(v)$ for function symbol $f\in\sigma'$, for $1\leq j<i\leq h$ and $\epsilon\in\{0,1\}$ expressing that $v$ is marked $P_i$ and that
$f(v)$ is marked $P_j$ (if $\epsilon=1$) or that $f^{-1}(v)$ contains a vertex marked $P_j$ (if $\epsilon=1$). Note that, in the later case, there is a single vertex marked $P_j$ in $f^{-1}(v)$, and that this vertex is the unique ancestor of $v$ marked $P_j$. Also note that $T_{f,i,j,\epsilon}(x)\rightarrow T_{E,i,j}(x)$ as all the functions of $\mathbf M$ follow adjacencies. It is easily checked that the interpretation $\mathsf I_Y$ constructs an encoding of $\mathbf M$ as a colored rooted tree with height at most $h$.

We now define the interpretation $\mathsf I_S$ that decodes 
a guided pointer structure (encoded in a rooted colored forest), as follows:
the adjacency relation $E$ is defined by a formula expressing that one of $x$ and $y$ (say $y$) is marked $P_i$, that the other (so, $x$) is the ancestor of $y$ marked $P_j$ (with $j<i$) and that $y$ is also marked $T_{E,j,i}$.
The functions $f\in\sigma'$ are analogously defined using the marks $T_{f,j,i,\epsilon}$.
Now, it is easily checked that
$\mathsf I_S(\mathsf I_Y(\mathbf M^+))=\mathbf M^+$.

\subsection{Modulo counting elimination in bounded height forests}

Let $Y$ be a (colored) rooted forest with height at most $h$, and let $\pi:V(Y)\rightarrow V(Y)$ be the mapping that maps each non-root vertex of $Y$ to its parent and each root vertex to itself. (Note that $\pi$ is definable as the height is bounded.)
For a vertex $v\in V(Y)$, we denote by
$Y\downarrow v$ the subtree of $Y$ induced by $v$ and its descendants, and rooted at $v$. For a tuple $\bar v$ of vertices, we denote by $Y\uparrow\bar v$ the subforest of $Y$ induced by all the vertices in the support of $\bar v$ and their ancestors. Note that the roots of $Y\uparrow\bar v$ are roots of $Y$.

It is known that, up to some linear-time computable monadic expansion, testing a formula on a bounded height forest reduces to checking whether the subforest induced by the free variables (and their ancestors) is label-isomorphic to some rooted forest in a finite set.
Formally, a \emph{$k$-labeled rooted colored forest} is a pair $(F,\lambda)$, where $F$  is a rooted colored forest and $\lambda: [k]\rightarrow V(F)$ is a mapping. The $k$-labeled rooted colored forest is \emph{tight} if every vertex of $F$ either is in the image of $\lambda$ or has a descendant in the image of $\lambda$.

Let $Y$ be a rooted colored forest, let $F$ be a tight $k$-labeled rooted colored forest, and let $\bar v$ be a $k$-tuple of vertices of~$Y$. Then, $\bar v$ \emph{defines an instance} of $F$ in $Y$ if the subforest $Y\uparrow \bar v$ induced by all the vertices in $\bar v$ and their ancestors in $Y$ is isomorphic to $F$ with the vertex $v_i$ corresponding to the vertex $\lambda(i)$ for $i=1,\dots,k$.
(This we denote $Y\uparrow {\bar v}\simeq (F,\lambda)$.)

A fundamental property of bounded height forests is the following (See, for instance, \cite{SBE_TOCL}): 
\begin{lemma}
	\label{lem:tree_elim}
	For every two integers $h,k$, and every first-order formula $\varsigma(\bar x)$ (in the language of colored graphs) with $|\bar x|=k$, there exist a linear-time computable function
	$\Lambda_E$ and a finite family $\mathcal F_\varsigma$ of 
	tight $k$-labeled rooted colored forests with height at most $h$ such that, for all rooted colored forests $Y$ with height at most $h$ and every $k$-tuple $\bar v$ of vertices of~$Y$, $\Lambda_E(Y)$ is a monadic expansion of $Y$ with
	\[
	Y\models \varsigma(\bar v)\quad\iff\quad \exists (F,\lambda)\in\mathcal F_\varsigma:\ \Lambda_E(Y)\uparrow{\bar v}\simeq (F,\lambda).
	\]
\end{lemma}

With \Cref{lem:tree_elim} in hand, we now show how to eliminate a modulo counting quantifier in classes of bounded height rooted forests.

\begin{lemma}
	\label{lem:count-elim}
	For every two integers $b,h$ and every first-order formula $\varsigma(\bar x,y)$ (in the language of colored graphs) there exists a linear-time computable map $L_Y$
	and first-order formulas $\zeta_{c,b}(\bar x)$  ($c\in\mathbb Z_b$)
	such that, for every rooted forest $Y$ with height at most $h$, $L_Y(Y)$ is a monadic expansion of $Y$ satisfying
	\[
	Y\models\mexists{c}{b}y\ \varsigma(\bar v,y)\quad\iff\quad
	Y^\ast\models\zeta_{c,b}(\bar v).
	\]
\end{lemma}

\begin{proof}
	According to \Cref{lem:tree_elim}, 
there exists a finite family $\mathcal F$ of tight
$k$-labeled rooted colored forests with height at most $h$ such that
$Y\models \varsigma(\bar v,w)$ if and only if there exists $(F,\lambda)\in\mathcal F$ with $\Lambda_E(Y)(\bar vw)\simeq (F,\lambda)$. Moreover, we may assume, without loss of generality, that some predicates $P_1,\dots,P_h$ encode the heights of the vertices of $Y$. In particular, $P_1(v)$ is equivalent to the property for $v$ of being a root of $Y$.

We consider two cases, depending on whether $\lambda(k+1)$ is in the same connected component of $F$ as some $\lambda(i)$ (with $i\in [k]$).

Assume $\lambda(k+1)$ is 
in a connected component of $F$ with no other image of $\lambda$. Let $F_1$ be the connected component of~$F$ containing $\lambda(k+1)$ and let $F_2=F\setminus F_1$.
We arbitrarily select a minimal subset $U$ of $[k]\times\{0,\dots,h\}$, such that each root of $F_2$ can be expressed as $\pi^j(x_i)$ for some $(i,j)\in U$. (Note that, by minimality, $|U|$ is the number of connected components of $F_2$.) 
At each root vertex $v$ of $Y$ we put a mark $B_i$, where $i$ is the number (modulo $b$) of copies of~$F_1$ in the connected component rooted at $v$, and we let $N\in\mathbb Z_b$ be the total count of these copies (modulo $b$). Note that the value~$N$ can be encoded using marks put on all the vertices of $Y$.
Note that all these copies indeed contain $v$ (as~$F_1$ contains a vertex marked $P_1$).  Let $\upsilon(\bar x)$ be a formula expressing
$Y\uparrow\bar x\simeq (F_2,\lambda)$. Then, the number (modulo $b$) of vertices $w$ such that $\Lambda_E(Y)\uparrow \bar vw$ is an instance of $(F,\lambda)$ is, modulo $b$, the difference between $n$ and the sum over $(i,j)\in U$ of the indices of the $B$-marks on $\pi^i(x_j)$. It follows that 
$\mexists{c}{b}w\in V(Y)\ \Lambda_E(Y)(\bar vw)\simeq (F,\lambda)$ can be expressed (in the monadic expansion of $\Lambda_E(Y)$) by means of a  first-order formula.

Otherwise, if $\lambda(k+1)$ is an ancestor of some $\lambda(i)$ with $i\in [k]$, then the number of instances $F$ in $\Lambda_E(Y)$ where $v_i=\lambda(i)$ for $i\in[k]$ is equal to the number of instances of $F_2$ in $\Lambda_E(Y)$ where $v_i=\lambda(i)$ for $i\in[k]$ (thus, this number is either $0$ or $1$).

Otherwise, let $s$ be the  highest ancestor of $\lambda(k+1)$ that is an ancestor of some $\lambda(i)$ for $i\in[k]$.
The forest $F$ splits into two parts, $F_1$ and $F_2$. The forest 
$F_1$ is the subtree of $F$ rooted at $s$, while $F_2$ is the subforest of $F$ obtained by deleting all the descendants of $s$ (distinct from $s$), as shown in Fig.~\ref{fig:pattern}.

\begin{figure}[ht]
	\begin{center}
		\includegraphics[width=.8\textwidth]{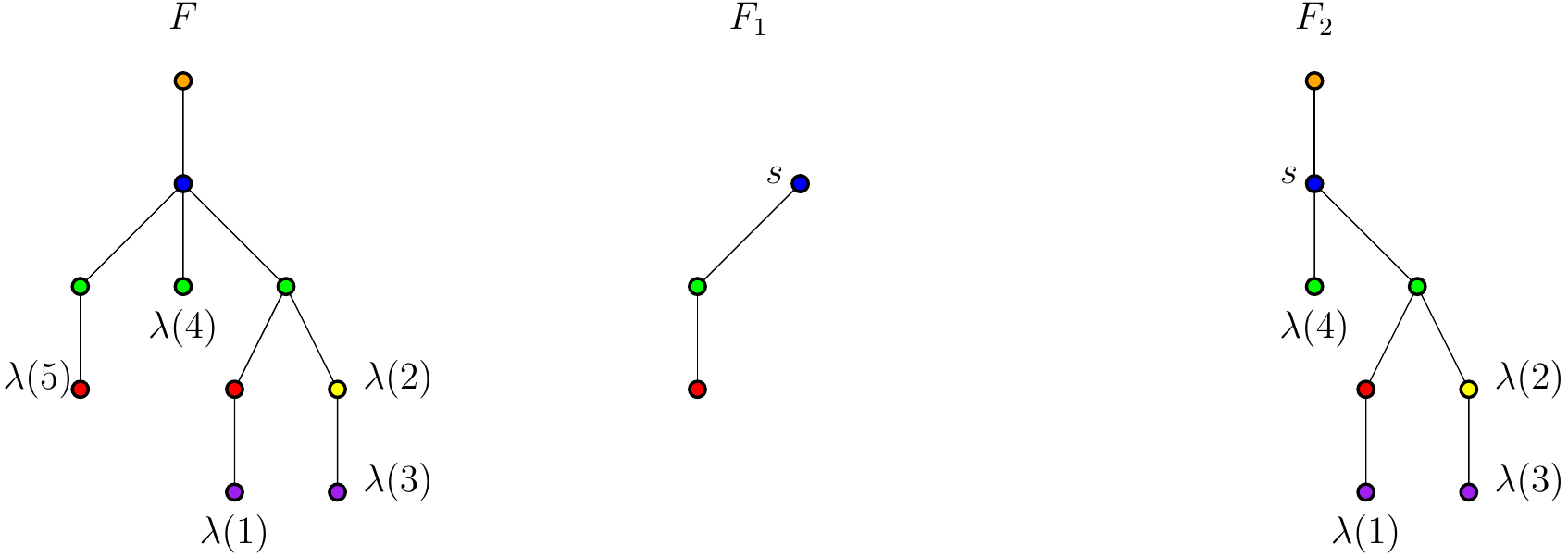}
		\caption{A connected pattern $F$ and its decomposition.}
		\label{fig:pattern}
	\end{center}
\end{figure}

The number of instances of $F$ in $\Lambda_E(Y)$ where $v_i=\lambda(i)$ for $i\in[k]$ is obtained as follows, by inclusion/exclusion: First, we consider the instance of $F_2$ in $\Lambda_E(Y)$ where $v_i=\lambda(i)$ for $i\in[k]$. (Note that this instance is unique, if it exists, as $F_2$ is tight.)
Let $v_s$ be the vertex of this instance corresponding to the vertex $s$ of $F_2$. 
From the number  of instances of $F_1$ rooted at $v_s$ we deduct the sum, over all the children $v'$ of $v_s$, of the number of instances of $F_1$ without its root rooted  at $v'$ (See Fig.~\ref{fig:count}).

\begin{figure}[ht]
	\begin{center}
		\includegraphics[width=.6\textwidth]{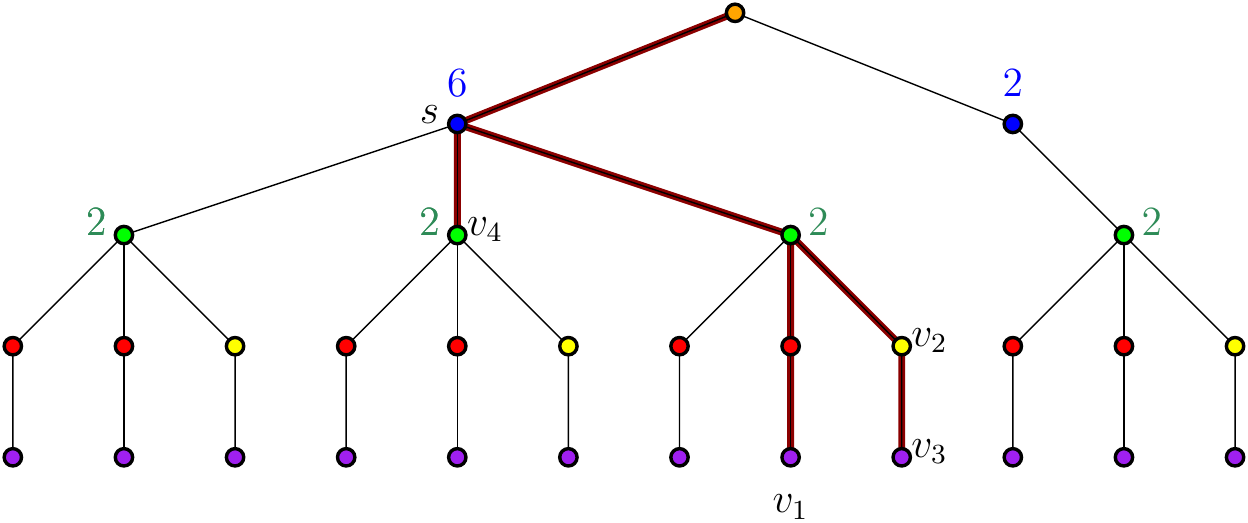}
		\caption{Counting. The blue values indicate the number of instances of $F_2$, while the green ones indicate the number of instances of $F_2$ without its root. Hence, the number of ways to augment the instance of $F_1$ (in bold) into an instance of $F_2$ is the blue label at $s$ (6) minus the sum of the green labels at children of $s$ in the instance of $F_2$ ($2+2$).
		The formula associated to a count of $2$ modulo $3$ could be
$
(\pi(\pi(v_1))=\pi(v_2))\wedge(\pi(v_3)=v_2)\wedge(\pi(v_4)=\pi(\pi(v_2)))\wedge
\bigvee_{i-(j+j)=2\mod 3} B_i(\pi(v_4))\wedge G_j(v_4)\wedge G_k(\pi(v_2))$, where $B_i$ stands for a blue count of $i\bmod 3$ and $G_j$ stands for a green count of $j\bmod 3$.}
		\label{fig:count}
	\end{center}
\end{figure}

By reducing these counts modulo $b$, we deduce that, in a proper monadic lif $Y^\ast$ of $\Lambda_E(Y)$, there are formulas $\zeta_{c,b}$ with
\[
Y\models\mexists{c}{b}y\ \varsigma(\bar v,y)\quad\iff\quad Y^\ast\models \zeta_{c,b}(\bar v).
\]
\qed
\end{proof}
\subsection{The main proof}
We prove \Cref{thm:main} in an inductive way.
\begin{proof}[of \Cref{thm:main}]
If $\varphi\in\FO[\sigma]$, then the result follows from \Cref{thm:DKT}.
If the statement holds for $\varphi_1$ and $\varphi_2$, it holds for any Boolean combination of these. Similarly, using the first case, if the statement holds for $\varphi$, it also holds for $\exists x\,\varphi$.
It follows that the only remaining case we have to consider is the case where $\varphi$ has the form
$\varphi(\bar x):=\mexists{a}{b}y\ \rho(\bar x,y)$, where $\rho$ is a quantifier-free formula.

The proof involves the transformation of a guided pointer structure $\mathbf M\in\mathscr C$ into a sequence of structures, eventually ending with a monadic expansion $L(\mathbf M)$ of $\mathbf M$ (See the proof scheme in \Cref{sec:overview}; we preserve the notations).
%
While describing these transformations, we also define new formulas, eventually ending with the definition of the quantifier-free formula $q$.
As we will consider formulas containing arbitrary compositions of functions, it will be practical to use the following convention. We assume that the functions in the signature $\sigma$ are indexed by integers, and are named $f_1,f_2,\dots$. Then, for a tuple 
$\bar\alpha$ with integer values, we denote by $f_{\bar\alpha}$ the composition\footnote{assuming all the $f_{\alpha_i}$ belong to $\sigma$}
\[
f_{\bar\alpha}=f_{\alpha_1}\circ\dots\circ f_{\alpha_{|\alpha|}}.
\]

As the formula $\rho$ is quantifier-free, the terms appearing in $\rho$ have all the form $f_{\bar\alpha}(x_i)$ or $f_{\bar\alpha}(y)$. 
Let $T$ be the set of all the tuples $\bar\alpha$ appearing in these expressions of the terms of $\rho$, let $k$ be the number of free variables of $\varphi$, and let $p=(k+1)\,|T|$.

According to \Cref{thm:chiBE_alg}, there exists an integer $N$ such that for every $\mathbf M\in\mathscr C$, a $(p+1)$-centered coloring $\gamma$ of $\Gaif(\mathbf M)$ with at most $N$ colors 
can be computed in linear time. Let $\gamma:M\rightarrow [N]$ be such a $(p+1)$-centered coloring and let $V_i(M)=\gamma^{-1}(i)$.  Adding $V_1,\dots,V_N$ as new unary relations, we get a linear-time computable monadic expansion $\mathbf M^+$ of $\mathbf M$.

We associate to each $v\in M$ its \emph{color type} $\tp(v)\in [N]^T$, which is a mapping from $T$ to $[N]$ defined by
$\tp(v)(\bar\alpha)=\gamma(f_{\bar\alpha}(v))$.
To each color type $t\in [N]^T$, we associate the formula
\[
\theta_t(x):=\bigwedge_{\bar{\alpha}\in T}V_{t(\bar{\alpha})}(f_{\bar{\alpha}}(x)),\]
 which tests whether $\tp(x)=t$.
For each tuple $\bar t\in ([N]^T)^k$ and each $t'\in [N]^T$ we define the formula
\[
\rho_{\bar t,t'}(\bar x,y):=\biggl(\bigwedge_{i\in [k]}\theta_{t_i}(x_i)\biggr)\wedge\theta_{t'}(y)\wedge\rho(\bar x,y).
\]

We also define the mapping $\mathcal P$, which maps  an element of $\mathbb Z_b$ to a set of functions from $[N]^T$ to $\mathbb Z_b$ as follows:
\[
\mathcal P(n):=\Bigl\{r\in\mathbb Z_b^{[N]^T}\Bigl| \sum_{t\in [N]^T} r(t)=n\Bigr\}.
\]

Then, we have the equivalence
\[
\mathbf M\models\mexists{a}{b}y\ \rho(\bar v,y)\qquad\iff\qquad \mathbf M^+\models\bigvee_{\bar t\in ([N]^T)^k}\quad\bigvee_{r\in\mathcal P(a)}\quad\bigwedge_{t'\in [N]^T}\quad 
\mexists{r(t')}{b}y\ \ \rho_{\bar t,t'}(\bar v,y).
\]

Let $C(\bar t,t')=\{t_i(\bar\alpha)\mid i\in [k],\bar\alpha\in T\}\cup\{t'(\bar{\alpha})\mid \bar{\alpha}\in T\}$ and
let $\mathbf M^+_{\bar t,t'}=\mathbf M^+[\bigcup_{i\in C(\bar t,t')}V_i]$.
Observe that 
$\mathbf M^+\models \rho_{\bar t,t'}(\bar v,w)$ if and only if
$\tp(v_i)=t_i$ (for all $i\in[k]$), $\tp(w)=t'$, and $\mathbf M_{\bar t,t'}^+\models \rho_{\bar t,t'}(\bar v,w)$. Thus, we have
\[
\mathbf M^+\models\mexists{c}{b}y\ \ \rho_{\bar t,t'}(\bar v,y)\quad\iff\quad \mathbf M_{\bar t,t'}^+\models\mexists{c}{b}y\ \ \rho_{\bar t,t'}(\bar v,y).
\]


As the coloring $\gamma$ is a $(p+1)$-centered coloring of $\Gaif(\mathbf M)$, and as $M^+_{\bar t,t'}$ is induced 
by the classes $V_i$ with $i\in C(\bar t,t')$, and as $|C(\bar t,t')|\leq (k+1)\,|T|=p$, we get ${\rm td}(\Gaif(\mathbf M^+_{\bar t,t'}))\leq p$. We define the rooted colored forests $Y^+_{\bar t,t'}$ as in \Cref{sec:Y}. In particular, $\mathbf M^+_{\bar t,t'}=\mathsf I_S(Y^+_{\bar t,t'})$. Hence,
\[
\mathbf M^+_{\bar t,t'}\models\mexists{c}{b}y\ \ \mathsf \rho_{\bar t,t'}(\bar v,y)
\quad\iff\quad Y^+_{\bar t,t'}\models\mexists{c}{b}y\ \ \mathsf I_S^\ast(\rho_{\bar t,t'})(\bar v,y).
\]

According to \Cref{lem:count-elim}, there exists a linear-time computable monadic expansion $Y^\ast_{\bar t,t'}$ of $Y^+_{\bar t;t'}$ and formulas $\zeta_{\bar t,t',c,b}(\bar x)$ such that 
\[
	Y^+_{\bar t,t'}\models\mexists{c}{b}y\ \ \mathsf I_S^\ast(\rho_{\bar t,t'})(\bar v,y)\quad\iff\quad Y^\ast_{\bar t,t'}\models \zeta_{\bar t,t',c,b}(\bar v).
\]

In the applications of  \Cref{lem:count-elim}, we keep the expanding  predicates distinct by further indexing them by
$\bar t,t'$. This will prevent clashes when copying all these predicates together.
Copying the monadic expansion  $Y^+_{\bar t,t'}\rightarrow Y^\ast_{\bar t,t'}$ to $M^+_{\bar t,t'}$, we get a monadic expansion  $M^\ast_{\bar t,t'}$ of  $M^+_{\bar t,t'}$, of which
 $Y^\ast_{\bar t,t'}$  is a transduction (by the transduction $\mathsf I_Y$). Thus, transferring $\zeta_{\bar t,t',c,b}(\bar v)$ we 
 get 
 \[
 Y^\ast_{\bar t,t'}\models \zeta_{\bar t,t',c,b}(\bar v)\quad\iff\quad \mathbf M^\ast_{\bar t,t'}\models\mathsf I_Y^\ast(\zeta_{\bar t,t',c,b})(\bar v).
 \]

   The monadic expansions $M^+_{\bar t,t'}\rightarrow M^\ast_{\bar t,t'}$ using relations
   indexed by $\bar t$ and $t'$, we can merge them safely and obtain a monadic expansion $\mathbf M^\ast$ of $\mathbf M^+$.
   Define
   \[
   \zeta(\bar x):=\bigvee_{\bar t\in ([N]^T)^k}\quad\bigvee_{r\in\mathcal P(a)}\quad\bigwedge_{t'\in [N]^T}\quad \mathsf I^\ast_Y(\zeta_{\bar t,t',r(t'),b})(\bar x).
   \]
   
Then, from what precedes, we get
   \[
   \mathbf M^\ast\models \zeta(\bar v)\quad\iff\quad \mathbf M\models\mexists{a}{b}y\,\rho(\bar v,y).
   \]
  
  According to \Cref{thm:DKT}, there exists an expansion $L(M)$ of $M^\ast_{\bar t,t'}$ 
  by some unary functions (following the adjacency)
  and a quantifier-free formula $q(\bar x)$ such that
  \[
  L( \mathbf M)\models q(\bar v)\quad\iff\quad  \mathbf M^\ast\models \zeta(\bar v).
  \]
\qed\end{proof}

\section{Shallow vertex minors}

The \emph{local complementation} of a graph $G$ at a vertex $v$ is the graph $G\oplus v$ obtained by complementing the adjacency relation between the neighbors of $v$ in $G$ \cite{bouchet1987isotropic}. A graph $H$ is a \emph{vertex minor} of a graph $G$ if it can be obtained from $G$ by a sequence of local complementations and vertex deletions \cite{OUM200579}. 
%
%

	A \emph{depth-$1$  vertex minor} of a graph $G$ is a graph $H$ obtained from $G$ by the local complementation of the vertices of an independent set $I$ of $G$ (in an arbitrary order), then the deletion of a subset $S$ of vertices. It is denoted as $G\oplus I-S$.
	Inductively, a \emph{depth-$k$  vertex minor} of a graph $G$ is defined as a depth-$1$ vertex minor of a depth-$(k-1)$ vertex minor of $G$.
Note that	 $H$ is a depth-$k$  minor of $G$ if and only if there exist subsets $I_1,\dots,I_k,S$ of $V(G)$ with the following properties:
	 \begin{enumerate}
	 	\item for each $0\leq i<k$, $I_{i+1}$ is an independent set of 
	 	$(\dots(G\oplus I_1)\oplus\dots)\oplus I_i$ (of $G$ if $i=0$);
	 	\item $H=((\dots(G\oplus I_1)\oplus\dots)\oplus I_k)-S$.
	 \end{enumerate}

The notion of vertex minor can be related to the model theoretical notion of NIP class:
A vertex-minor closed class of graphs is NIP if and only if it has bounded cliquewidth.
Indeed, if a class {\Cc} includes all circle graphs, then {\Cc} has unbounded cliquewidth and is easily seen not to be  NIP. Otherwise, the class {\Cc} has bounded cliquewidth \cite{GEELEN2020}, thus is~NIP.

Excluding given vertex minors at each depth, in the spirit of nowhere dense classes, we get more general classes of graphs. For instance, as bounded depth vertex minors can be constructed by means of \FOM-transductions, we deduce from
the results of \cite{bonnet2022twin} that bounded depth vertex minors of graphs with bounded twin-width have bounded twin-width.

From \cite{SBE_TOCL} it follows that every structurally bounded expansion class can be obtained as depth-$1$ vertex minors of graphs in a class with bounded expansion. Conversely, as bounded depth vertex minors can be obtained by {\FOM} transductions, we deduce from  \Cref{cor:trans} that bounded depth vertex minors of graphs in a structurally bounded expansion class form a structurally bounded expansion class. Thus, we have the next corollary.

\begin{corollary}
	\label{cor:vm}
	A class $\mathscr C$ has structurally bounded expansion if and only if it there is a bounded expansion class $\mathscr D$ and an integer $d$ such that every graph in $\mathscr C$ is a depth-$d$ vertex minors of a graph in $\mathscr D$.
\end{corollary}

%

%

%

\section{Matrices over finite rings}

The multiplication of two $n\times n$ matrices is one of the most basic algebraic problems, and considerable effort was
devoted to designing efficient algorithms to compute it. 
While the naive algorithm performs $O(n^3)$ operations, the algorithm designed by Alman and Williams~\cite{alman2021refined} needs only $O(n^{2.37188})$ operations. When the matrices are sparse, further improvements can be done (See~\cite{yuster2005fast}).

It has been recently proved that matrix multiplications can be computed 
efficiently on any subclass of ${\rm Mat}_n(\mathbb F_p)$ with bounded twin-width~\cite{bonnet2022twin}. The results we present here share two ideas with the ones obtained for multiplications of matrices with small twin-width. First, we use \FOM-interpretation to perform the computation, which relies on the above fast \FOM-model checking algorithm. Second, we can use an adapted data-structure to represent both our input matrices and our result in $O(n)$-space, allowing $O(n)$-time computations, and constant-time 
 to return the value of an arbitrary entry.

Standard computations on matrices we shall consider include addition, multiplication, transposition, scalar multiplication. 
Note that if $R$ is a simple finite ring,
then computations can be reduced to computations on matrices over a finite field $\mathbb F_q$,
 as any finite simple ring is isomorphic to the ring 
${\rm Mat}_n(\mathbb{F}_q)$ for some $n$ and $q$.

\begin{definition}
	A class $\mathcal M\subseteq \bigcup_{n\geq 1}{\rm Mat}_n(R)$ of matrices  over a finite ring $R$ has \emph{bounded expansion support} if the class $\Supp(\mathcal M)$ of all graphs
	$G([n],E)$ where $ij\in E$ if there exists an $n\times n$ matrix $M\in\mathcal M$ with $M_{i,j}\neq 0$ or $M_{j,i}\neq 0$ has bounded expansion.
\end{definition}

	A class $\mathcal M\subseteq {\rm Mat}_n(\mathbb{F}_p)$ of matrices  over a field $\mathbb F_p$ has \emph{bounded rank} is there exists an integer $r$ such that every matrix $M\in\mathcal M$ has rank at most $r$ over $\mathbb F_p$. 
	In the case where the entries are not in a field, it is more difficult to define a notion of rank. We shall use the following notions:

		Let $M$ be an $n\times n$ matrix. 
		The \emph{domain} of $M$ is the set $D$ of all the non-zero elements arising as an entry of $M$.
		For an element $d\in D$, the \emph{$d$-slice} of $M$ is the matrix $M(d)$, whose $ij$ entry is $1$ if $M_{i,j}=d$, and $0$ otherwise.
		Note that $M=\sum_{d\in D}d\,M(d)$.
		
		The \emph{set-rank} of $M$
		is defined by
\[
	\srank(M)=:\sum_{d\in D}\rank_2 M(d),
\]
where $\rank_q$ denotes the rank over the field $\mathbb F_q$.
Note that if $R=\mathbb F_p$ is a field, then
$\rank_p(M)\leq \srank(M)\leq p\,\rank_p(M)$.

\begin{lemma}
	Let $R$ be a finite ring with $p$ elements, let $r$ be a positive integer, let $W_{d,\ell}$ and $V_{d,\ell}$ with
	$d\in [p],\ell\subseteq [r]$ be a set of $p2^r$ unary predicates (called marks).
For $d\in[p]$, let
	$q_d(x,y)$ be the first-order formula defined by 
\[
	q_d(x,y):=\bigvee_{\ell\subseteq[r]} W_{d,\ell}(x) V_{d,\ell}(y).
\]

	 Then, for every integer $n$ and every $M\in {\rm Mat}_n(R)$  with set-rank at most $r$, there exist 
	 a marking $[n]^*$ of $[n]$ with marks $W_{d,\ell}$ and $V_{d,\ell}$,
	 such that for every $d\in[p]$ we have
\[
	M_{i,j}=d\quad\iff\quad [n]^*\models q_d(i,j).
\]
	 
\end{lemma}
\begin{proof}
	For each integer $d$, let $A^{d,1},\dots,A^{d,r_d}$ be a basis of the line vectors of $M(d)$ over $\mathbb F_2$, 
	Then, every line  $L_i$ of $M(d)$ is equal the sum of some of the $A^{d,j}$'s. We mark $i$ by $W_{d,\ell}$ if 	$L_i=\sum_{i\in\ell} A^{d,i}$. 
	Moreover, we mark $j$ by $V_{d,\ell}$ if 
	the $j$th entry of $\sum_{k\in\ell}A^{d,k}$ is $1$. Then, $M_{i,j}=d$ if and only if, for some $\ell\subseteq [r]$, $i$ is marked by $W_{d,\ell}$ and $j$ is marked by $V_{d,\ell}$.
	\qed
\end{proof}

\begin{corollary}
	\label{thm:mat-calc}
	Let $R$ be a finite ring, let $\mathcal M\subseteq {\rm Mat}_n(R)$ be a class of matrices with bounded expansion support, and let $\mathcal R\subseteq {\rm Mat}_n(R)$ be a class of matrices with bounded set-rank (playing the role of constants).
	
	Then, for every function $F:\mathcal M^a\times\mathcal R^b\rightarrow {\rm Mat}_n(R)$ which is a composition of matrix additions, products, transpositions,  Hadamard products, and scalar products, and using the $n\times n$ identity matrix $I_n$ (as a constant), there is a quadratic time algorithm to compute $F$. (Or even linear time if we work with appropriate representations.)
\end{corollary}

\begin{proof}[Sketch]
We encode all the matrices used by $F$ in a single graph as a vertex- and edge-colored directed graph (with no parallel arcs). Then, to the function $F$ we associate an \FOM-formula
$\psi_F$, such that the result of $F$ will correspond to the interpretation of the colored directed graph using $\phi_F$. Then, $\phi_F$ is equivalent to an existential formula~$\phi_F$ in an expansion (that can be computed in linear time). Using a low tree-depth decomposition, this existential formula is equivalent to some  property that can be checked in constant time  in a monadic expansion (computable in linear time) of the tree models of the low tree-deph decomposition. 
\qed
\end{proof}
It is interesting to note that, for this proof,
 a weaker assumption suffices. We need that the union of the supports of any $a$ matrices in $\mathcal M$ is included in the edge set of some graph in a bounded expansion class (for some appropriate ordering of the vertex set).

\section{Discussion}

In view of the obtained results, it is not clear that \FOM-transductions are, in general, more powerful than \FO-transductions. This leads to the following conjecture and problem.

\begin{conjecture}
	Let $\mathscr C$ and $\mathscr D$ be infinite classes of finite graphs.
	Then, $\mathscr D$ is an \FOM-transduction of $\mathscr C$ if and only if $\mathscr D$ is an \FO-transduction of $\mathscr C$.
\end{conjecture}





\bibliographystyle{elsarticle-num-names-alpha}
\bibliography{ref}







\end{document}